# Spacetime-topological events

**Authors:** Joshua Feis[1,2]†, Sebastian Weidemann[1]†, Tom Sheppard[3], Hannah M. Price[3], Alexander Szameit[1]*

**Affiliations:**

[1]Institute of Physics, University of Rostock, Rostock, Germany.

[2]Department of Engineering Science, University of Oxford, Oxford, United Kingdom.

[3]School of Physics and Astronomy, University of Birmingham, Birmingham, United Kingdom.

*Corresponding author. Email: alexander.szameit@uni-rostock.de

†These authors contributed equally to this work.

**Abstract:** Time is, figuratively and literally, becoming the new dimension for crystalline matter. As such, rapid recent progress on time-varying media gave rise to the notion of temporal and spatiotemporal crystals. Fundamentally rethinking the role of time, which, in contrast to space exhibits a unique unidirectionality often referred to as the "arrow of time", promises a new dimension also for topological physics. Here, we enter the new realm of time and spacetime topology: Firstly, we implement a time-topological time interface state. Secondly, we propose and observe a spacetime-topological event and demonstrate unique features like its limited collapse under disorder and causality-suppressed coupling. The new paradigms of time and spacetime topology unveil a distinctive role of causality and non-Hermiticity in topology and pave the way towards topological spatiotemporal wave control with unique robustness.

**Main Text:** Crystals are matter with the highest degree of spatial order, with their structure being arranged in spatially repeating patterns. Recently, this pillar of science has been fundamentally rethought (*1*). As spatial crystals are defined by their periodicity in space, the far newer concepts of temporal (*2–4*) and spatiotemporal (*5, 6*) crystals refer to structures that repeat periodically in time and spacetime, respectively. In photonics specifically, temporal crystals are also referred to as time-varying media or photonic time crystals (*2*) and they are reminiscent of Wilczek's proposal of time crystals (*1, 7*): In time crystals specifically, continuous time-translation symmetry is broken spontaneously, often as a result of (quantum) many-body interactions, whereas temporal crystals, as we consider them here, generally refer to any system with discrete time-translation symmetry only. Rethinking the role of time promises a new dimension also for the celebrated field of topological physics, setting paradigms of time and spacetime topology: Topological physics has become a unifying theme for diverse phenomena from seemingly distant fields, ranging from the celebrated quantum Hall effects for electrons (*8, 9*), to equatorial ocean waves (*10*) and with promising applications like topological quantum computation (*11*) based on the world line braiding of anyons (*12*), or topological lasers (*13–15*) enabled by photonic modes that are intrinsically resistant to scattering (*16*). Yet, these topological features have so far relied on energy gap topology at spatial boundaries only, i.e., the emergence of robust, long-lived states residing within an energy gap while being localized at a spatial topological interface (Fig. 1A), which is the change of the relevant topological invariant in space (*17, 18*).

However, time is a new dimension for topology with fundamental differences to its spatial counterpart: Its unique unidirectionality, often called the arrow of time (*19*), endows time with an intrinsic asymmetry which space does not exhibit. This asymmetry leads to strikingly different physics, for instance, such as time not supporting back-reflections but instead, against all intuition, mandates the emergence of a unique notion of time reflections (*20*). Correspondingly, unlike the familiar space topology, time topology may lead to topologically protected states that are localized at topological temporal interfaces (Fig. 1B, bottom), which are formed by the temporal change of a time-topological invariant (*21*). It is based not on energy but momentum gaps (*21, 22*) (Fig. 1B), i.e., regions of purely imaginary energies, which can generically appear in driven and dissipative (non-Hermitian (*23, 24*)) systems, implying an intrinsic connection between time topology and non-Hermiticity, unlike space topology for which non-Hermiticity is not generally necessary.

Here, we demonstrate time-topological states localized at the temporal boundary of two photonic lattices with different momentum gap topology. We further introduce a time-topological invariant and establish a relation between this invariant and the observed time-topological states using a transfer matrix approach. Our experiments are based on a photonic lattice implemented through light propagation in coupled optical fiber loops (*25–27*). The capability to modulate the optical properties of the fiber loops in time allows to open and close momentum and energy gaps and to control their topology. Transcending the separate concepts of space and time topology, we introduce the notion of spacetime topology by observing a spacetime-topological event, i.e., where a topological state localizes at a single point in spacetime (Fig. 1C, bottom). The topological event emerges at the intersection of energy-momentum gapped photonic lattices whose topology is determined by a spacetime-topological invariant that we introduce. Importantly, this topological state is localized in all available dimensions including time, thus necessarily requiring the interplay of space and time in any causal system.

*Theory.* We start by discussing the most fundamental scenario, i.e., a one-dimensional two-band model, whose energy and momentum gaps can be independently controlled. In a second step, we deliberately close and open momentum and energy gaps to control their corresponding topology. This allows our model to host topological states at temporal interfaces, which spectrally reside in the momentum gap, but also the well-established space topological counterparts, which are instead localized at spatial interfaces and reside in the energy gap. Finally, we realize a combined energy-momentum gap that features topological states that are localized at a single point in (1+1)D spacetime, and as such constituting a topological event.

Our model is a time-varying implementation of the archetypal Su-Schrieffer-Heeger (SSH) model (*28*) with an additional gain-loss modulation that corresponds to a time-periodic drive of an imaginary on-site potential. The resulting evolution can be described in the framework of discrete-time quantum walks of single particles, as detailed in Section 3 of the Supplementary Text, and then implemented via the light propagation in coupled optical fiber loops as shown in Fig. 2. The experimental implementation employs a temporal encoding of pulse arrival times (*29, 30*), creating a coupling of light between time-bins that yields the dynamics of a photonic mesh lattice with a synthetic time (exhibiting the characteristic unidirectionality of time) and space dimension, as detailed in Section 2 of the Supplementary Text.

The quantum walk describes the evolution of a quantum particle on discrete lattice positions $x$, which are coupled to neighbouring positions in a stepwise protocol along the time axis $t$. The dynamics are governed by the recursive evolution equations

$$\begin{aligned} u_x^{t+1} &= [\cos(\beta)\, u_{x+1}^t + i\sin(\beta)\, v_{x+1}^t] e^{i\varphi_u} \\ v_x^{t+1} &= [i\sin(\beta)\, u_{x-1}^t + \cos(\beta)\, v_{x-1}^t] e^{i\varphi_v}, \end{aligned} \qquad (1)$$

where $u_x^t$ and $v_x^t$ are the amplitudes of the spin-like degree of freedom at lattice point $x$ and time step $t$, corresponding to left and right-moving paths in the photonic lattice (Fig. 2B), respectively. The parameter $\beta(t,x)$ characterises the splitting ratio of the beamsplitters and, thus, mediates the coupling between lattice positions and spins. The lattice parameters $\text{Re}(\varphi_v)$ and $\text{Im}(\varphi_v)$ are additional real and imaginary phase contributions for each component $v$ picked up during evolution. By appropriately choosing $\varphi_v(t,x)$ as a function of spin $v \in \{u, v\}$, lattice position $x$ and time $t$, all three types of topological band gaps may be obtained. The resulting photonic lattice is shown in Fig. 2: The coupling alternates between $\beta_1$ and $\beta_2$ in a two-time-step fashion, thereby realising the SSH model, being accompanied by a four-time-step gain-loss modulation $\varphi_u(t+4) = \varphi_u(t)$ with $[\varphi_u(0), \varphi_u(1), \varphi_u(2), \varphi_u(3)] = [ig, 0, 0, -ig]$, where $g$ is the real gain-loss strength, which is antisymmetrically distributed on the spin-degree of freedom, i.e., $\varphi_v = -\varphi_u$.

The resulting energy spectra are shown in the bottom row of Fig. 1. Their calculation is detailed in Section 4 and 5 of the Supplementary Text. For $g = 0$, one obtains the Hermitian Floquet SSH model (*31*), which is gapped in energy (Fig. 1A). These energy gaps are ranges where the energy is real whereas the momentum is imaginary, leading to spatially decaying solutions called evanescent waves (*32*). Note, that in a crystal, momentum becomes quasimomentum due to discrete spatial translation symmetry (*33*), whereas here, additionally, energy likewise becomes quasienergy due to the discrete translation symmetry in time, as stated by Floquet theory (*34–36*). As a result, energies are $2\pi$-periodic and therefore a second energy gap at $E = \pm\pi$ may exist. For $g \neq 0$, i.e., with the addition of non-Hermiticity, and $\beta_1 = \beta_2 = \pi/4$, there are no energy but only momentum gaps (Fig. 1B), i.e., regions where the momentum is real whereas the energy is imaginary, leading to temporally growing and decaying solutions (*2*). Note, that in Floquet systems, such as the one considered here, the momentum gap condition also allows the real part of the energy to be $\pm\pi$. Finally, for $g \neq 0$ and $\beta_1 \neq \beta_2$, one can find an energy-momentum gap (Fig. 1c), where neither energy nor momentum are real (*6*).

The occurrence of gaps in the spectrum of allowed energies plays a major role in topological physics (*17*), as showcased by the celebrated topological states of the SSH model, which may also emerge here: Based on the eigenstates of the Bloch Hamiltonian, one can calculate a topological invariant, i.e., an integer winding number that remains invariant under perturbations that admit certain symmetries (*18*). Its calculation is based on integrating the phase of the eigenstates accumulated along the energy bands, as detailed in Section 6 of the Supplementary Text. The so-called bulk-boundary correspondence (*37*, *38*) then assures the emergence of topological states localized at spatial topological interfaces, which are defined as a spatial change of this (spatial) winding number. Note, that the winding number can only change upon closing the energy gap, which can here be achieved by altering the couplings $\beta_{1,2}$. In the following we will show that in the presence of momentum gaps, the topological description can be extended to enable the prediction of time-topological states as well as spacetime-topological states.

*Results.* As a starting point, with the Hermitian Floquet SSH model, we experimentally realize a photonic lattice that is gapped in energy and verify the emergence of spatial topological states. If the spatial winding number changes across the spatial interface one expects topological states localized at the interface. Our experiments agree with prior work (*31*, *39*), showing the emergence of a spatially localized state only when the aforementioned difference is nonzero (Fig. 3A, bottom). Conversely, for the trivial case of equal winding numbers across the interface, no topological state can be seen (Fig. 3A, top).

However, this situation changes when closing the energy gap while opening a momentum gap (Fig. 1B, center) to approach the realm of time topology. Experimentally, we achieve this by matching the two couplings $\beta_1 = \beta_2$ and applying a gain-loss modulation with $g = 0.1$, thus turning the system non-Hermitian. Importantly, its topological characterisation must now be rethought: On the one hand, since the energy gap is now closed, the spatial winding number becomes ill-defined. On the other hand, though, there are now two momentum bands separated by the momentum gap. Momentum bands are potentially complex functions of real energies, like energy bands are potentially complex functions of real momenta, as shown in Fig. 1. Their calculation is detailed in the Section 5 of the Supplementary Text. The eigenstates of the system may accumulate a geometric phase along these momentum bands, motivating the definition of a temporal winding number. The calculation of the temporal winding number is detailed in Section 6 of the Supplementary Text. Moreover, using a transfer matrix approach, we show that the occurrence of time-topological states localized at temporal interfaces is predicted by changes of the temporal winding number in time (see Sections 7 and 8 of the Supplementary Text). Such a time-topological interface state can clearly be observed in our experimental results (Fig. 3b, bottom) showing an exponential localization of light that is centered on the temporal boundary across which the invariant changes. Note, that the inherent growth and decay of the time-topological state points at the non-Hermitian nature of time topology. It is, however, distinct from conventional non-Hermitian topology as the time-topological invariant naturally arises from the consideration of momentum and not energy bands. Importantly, this new type of topology is markedly different from concepts, in which time is treated as a non-causal synthetic spatial dimension (*40*), and correspondingly fall into the category of traditional space topology without the unique aspects that arise from causality and momentum gap topology. We also note, that related preliminary simulation results have indicated the existence time-topological states not just in temporal but also in spatiotemporal crystals (*41*). An analytical derivation of the shape of the time-topological interface state matching our experimental observations may be achieved by considering the projections of the left and right eigenstates before and after the interface (see Supplementary Text Sections 9 and 10). Conversely, for the trivial case of equal temporal winding numbers across the temporal interface, no such state can be seen (Fig. 3B, top).

In the final step of our work, we are investigating a possible intertwinement of the familiar space topology with the notion of time topology introduced here. To this end, we probe whether a notion of spacetime topology emerges in the presence of energy-momentum gaps. Based on the following results, we answer this question in the affirmative. By considering the product of the changes of the spatial and temporal winding number across their respective interfaces, we define a spacetime-topological invariant for systems with combined energy-momentum gaps (see Supplementary Text Section 6). This invariant predicts the emergence of spacetime-topological events, i.e., where a topological state is exponentially localized on a single point in spacetime. At this point, spacetime regions, which are separated by a spatial and a temporal interface with different respective invariants across them, meet. We experimentally probe the light dynamics in the spacetime-topological and trivial case as shown in Fig. 4. In the trivial case, the light evolution neither shows the distinct exponential spatial localization nor any temporal localization (Fig. 4A). In the topological case, the topological invariant correctly predicts the emergence of a light field that is exponentially localized along the time and space axis, while being located exactly at the designated interface crossing, thus forming the spacetime-topological event (Fig. 4B).

This intertwinement of spacetime with topology gives rise to unique effects with no analogue in traditional topology: For instance, causality can suppress coupling into the topological event even when the excitation has a non-zero overlap with it. If an excitation is within its past light cone (Fig. 5A, left), the spacetime-topological state will be excited when there is any overlap between the excitation and the profile of the state. However, if the excitation is not within the past light cone (Fig. 5A, right) of the spacetime-topological event, the topological state will not be populated no matter how large the overlap between the excitation and the state is. This causality-suppressed coupling is a direct result of the arrow of time: Even though the overlap is approximately the same in both cases, causality dictates that energy cannot couple into the topological event from its future but only from its past light cone, enabling a selective excitation and hence endowing these states with an additional robustness against stray excitation unique among topological states. In contrast, in space topology energy anywhere in the system will typically couple into and between topological states after some time, i.e., impeding ideal topological protection and reducing their robustness in finite systems.

Furthermore, spacetime-topological events exhibit unique behavior under disorder: Due to its distinct nature, it is possible for the energy-momentum gap to close only partially under disorder, leaving behind a momentum gap which maintains the temporal localization. This is shown in Fig. 5b, where it can be seen that for moderate disorder (Fig. 5B, left) the spacetime localization remains robust with only small distortions, whereas for strong disorder (Fig. 5B, right), the spatial localization collapses. However, this collapse is indeed limited to the spatial part of the spacetime localization only with the state remaining temporally localized. Such a limited collapse is markedly different from conventional topology, where closing the energy gap generally fully collapses all topological states, pointing at a new dimension of topological robustness where the collapse of topological localization may be limited to some but not all dimensions.

*Conclusion.* We have experimentally demonstrated time-topological states localized at temporal boundaries based on the time topology of the momentum gap of a photonic lattice. The time topological states are predicted by a momentum band-based winding number we proposed. Due to the growing and decaying nature of such states, their description naturally includes non-Hermiticity. We then transcend the separate notions of energy and momentum gap topology by demonstrating spacetime-topological events, where topological states localize at a zero-dimensional spacetime boundary. These topological events are predicted by a spacetime-topological invariant we defined. Their localization in all available dimensions including time necessitates a notion of spacetime topology, bringing forth unique features such as causality-suppressed coupling or the limited collapse of event states. Our results set the new paradigms of time and spacetime topology, based on the interconnections between the celebrated energy gap topology and introduced notion of momentum gap topology. As such, we anticipate our studies to have far-reaching implications: For the field of topological physics specifically, our work shows a new avenue away from the conventional sole focus on spatial features of topological systems, towards the wider scope of spacetime topology. The evident connection of time and spacetime topology to non-Hermiticity also stimulates the possibility of a connection to the thriving field of non-Hermitian topology. Moreover, our work invites the exploration of how the concepts of time and spacetime topology newly connect topological physics to other fields where the arrow of time plays an important role, such as thermodynamics. More generally, the ongoing efforts to engineer wave systems capable of ultrafast temporal modulations, with recent successes on implementing time interfaces in nonlinear optical media (*42*), ultracold atoms (*43*) as well as for water waves (*20*), point towards a wide range of potential future platforms for time- and spacetime-topological effects. In this vein, controlling momentum and energy gap topology may enable the topological shaping of various types of waves in space and time, with applications for example in spatiotemporal wave control for imaging or communication (*44*, *45*) as well as topological lasers (*46*).

**References and Notes**

1. F. Wilczek, Crystals in time. *Scientific American* **321**, 28–36 (2019).

2. E. Galiffi, R. Tirole, S. Yin, H. Li, S. Vezzoli, P. A. Huidobro, M. G. Silveirinha, R. Sapienza, A. Alù, J. B. Pendry, Photonics of time-varying media. *Adv. Photon.* **4** (2022).

3. N. Engheta, Four-dimensional optics using time-varying metamaterials. *Science* **379**, 1190–1191 (2023).

4. X. Wang, M. S. Mirmoosa, V. S. Asadchy, C. Rockstuhl, S. Fan, S. A. Tretyakov, Metasurface-based realization of photonic time crystals. *Science Advances* **9**, eadg7541 (2023).

5. C. Caloz, Z.-L. Deck-Léger, Spacetime metamaterials—part I: general concepts. *IEEE Transactions on Antennas and Propagation* **68**, 1569–1582 (2019).

6. Y. Sharabi, A. Dikopoltsev, E. Lustig, Y. Lumer, M. Segev, Spatiotemporal photonic crystals. *Optica* **9**, 585–592 (2022).

7. F. Wilczek, Quantum time crystals. *Physical review letters* **109**, 160401 (2012).

8. D. J. Thouless, M. Kohmoto, M. P. Nightingale, M. den Nijs, Quantized Hall Conductance in a Two-Dimensional Periodic Potential. *Phys. Rev. Lett.* **49**, 405–408 (1982).

9. J. P. Eisenstein, H. L. Störmer, The Fractional Quantum Hall Effect. *Science* **248**, 1510–1516 (1990).

10. P. Delplace, J. B. Marston, A. Venaille, Topological origin of equatorial waves. *Science* **358**, 1075–1077 (2017).

11. C. Nayak, S. H. Simon, A. Stern, M. Freedman, S. Das Sarma, Non-Abelian anyons and topological quantum computation. *Rev. Mod. Phys.* **80**, 1083–1159 (2008).

12. J. Nakamura, S. Liang, G. C. Gardner, M. J. Manfra, Direct observation of anyonic braiding statistics. *Nat. Phys.* **16**, 931–936 (2020).

13. M. A. Bandres, S. Wittek, G. Harari, M. Parto, J. Ren, M. Segev, D. N. Christodoulides, M. Khajavikhan, Topological insulator laser: Experiments. *Science* **359**, eaar4005 (2018).

14. B. Bahari, A. Ndao, F. Vallini, A. El Amili, Y. Fainman, B. Kanté, Nonreciprocal lasing in topological cavities of arbitrary geometries. *Science* **358**, 636–640 (2017).

15. P. St-Jean, V. Goblot, E. Galopin, A. Lemaître, T. Ozawa, L. Le Gratiet, I. Sagnes, J. Bloch, A. Amo, Lasing in topological edge states of a one-dimensional lattice. *Nature Photonics* **11**, 651–656 (2017).

16. T. Ozawa, H. M. Price, A. Amo, N. Goldman, M. Hafezi, L. Lu, M. C. Rechtsman, D. Schuster, J. Simon, O. Zilberberg, I. Carusotto, Topological photonics. *Rev. Mod. Phys.* **91**, 015006 (2019).

17. M. Z. Hasan, C. L. Kane, Colloquium : Topological insulators. *Rev. Mod. Phys.* **82**, 3045–3067 (2010).

18. J. K. Asbóth, L. Oroszlány, A. Pályi, *A Short Course on Topological Insulators* (Springer International Publishing, Cham, 2016; http://link.springer.com/10.1007/978-3-319-25607-8)vol. 919 of *Lecture Notes in Physics*.

19. A. S. Eddington, *The Nature of the Physical World: Gifford Lectures of 1927: An Annotated Edition* (Cambridge Scholars Publishing, Newcastle upon Tyne, 2014).

20. V. Bacot, M. Labousse, A. Eddi, M. Fink, E. Fort, Time reversal and holography with spacetime transformations. *Nature Physics* **12**, 972–977 (2016).

21. E. Lustig, Y. Sharabi, M. Segev, Topological aspects of photonic time crystals. *Optica* **5**, 1390 (2018).

22. J. Reyes-Ayona, P. Halevi, Observation of genuine wave vector (k or β) gap in a dynamic transmission line and temporal photonic crystals. *Applied Physics Letters* **107** (2015).

23. Y. Ashida, Z. Gong, M. Ueda, Non-hermitian physics. *Advances in Physics* **69**, 249–435 (2020).

24. R. El-Ganainy, K. G. Makris, M. Khajavikhan, Z. H. Musslimani, S. Rotter, D. N. Christodoulides, Non-Hermitian physics and PT symmetry. *Nature Physics* **14**, 11–19 (2018).

25. A. Regensburger, C. Bersch, M.-A. Miri, G. Onishchukov, D. N. Christodoulides, U. Peschel, Parity–time synthetic photonic lattices. *Nature* **488**, 167–171 (2012).

26. S. Weidemann, M. Kremer, T. Helbig, T. Hofmann, A. Stegmaier, M. Greiter, R. Thomale, A. Szameit, Topological funneling of light. *Science* **368**, 311–314 (2020).

27. S. Weidemann, M. Kremer, S. Longhi, A. Szameit, Topological triple phase transition in non-Hermitian Floquet quasicrystals. *Nature* **601**, 354–359 (2022).

28. W. P. Su, J. R. Schrieffer, A. J. Heeger, Solitons in Polyacetylene. *Phys. Rev. Lett.* **42**, 1698–1701 (1979).

29. T. Ozawa, H. M. Price, Topological quantum matter in synthetic dimensions. *Nature Reviews Physics* **1**, 349–357 (2019).

30. L. Yuan, Q. Lin, M. Xiao, S. Fan, Synthetic dimension in photonics. *Optica* **5**, 1396–1405 (2018).

31. T. Kitagawa, M. A. Broome, A. Fedrizzi, M. S. Rudner, E. Berg, I. Kassal, A. Aspuru-Guzik, E. Demler, A. G. White, Observation of topologically protected bound states in photonic quantum walks. *Nature communications* **3**, 882 (2012).

32. M. Born, E. Wolf, *Principles of Optics: Electromagnetic Theory of Propagation, Interference and Diffraction of Light* (Cambridge University Press, Cambridge, 1999).

33. N. W. Ashcroft, N. D. Mermin, *Solid State Physics* (Holt, Rinehart and Winston, New York, 1976).

34. M. Holthaus, Floquet engineering with quasienergy bands of periodically driven optical lattices. *Journal of Physics B: Atomic, Molecular and Optical Physics* **49**, 013001 (2015).

35. S. Yin, E. Galiffi, A. Alù, Floquet metamaterials. *ELight* **2**, 1–13 (2022).

36. M. S. Rudner, N. H. Lindner, Band structure engineering and non-equilibrium dynamics in Floquet topological insulators. *Nature reviews physics* **2**, 229–244 (2020).

37. C. Cedzich, F. A. Grünbaum, C. Stahl, L. Velázquez, A. H. Werner, R. F. Werner, Bulk-edge correspondence of one-dimensional quantum walks. *Journal of Physics A: Mathematical and Theoretical* **49**, 21LT01 (2016).

38. Y. Tanaka, A constructive approach to topological invariants for one-dimensional strictly local operators. *Journal of Mathematical Analysis and Applications* **500**, 125072 (2021).

39. N. Malkova, I. Hromada, X. Wang, G. Bryant, Z. Chen, Observation of optical Shockley-like surface states in photonic superlattices. *Optics letters* **34**, 1633–1635 (2009).

40. G. Žlabys, C. Fan, E. Anisimovas, K. Sacha, Six-dimensional time-space crystalline structures. *Phys. Rev. B* **103**, L100301 (2021).

41. O. Segal, E. Lustig, Y. Sharabi, M.-I. Cohen, R. Ziv, M. Lyubarov, A. Dikopoltsev, M. Segev, "Topology in Photonic Space-Time Crystals" in *Conference on Lasers and Electro-Optics* (Optica Publishing Group, San Jose, California, 2022; https://opg.optica.org/abstract.cfm?URI=CLEO_AT-2022-JW4A.4), p. JW4A.4.

42. Y. Zhou, M. Z. Alam, M. Karimi, J. Upham, O. Reshef, C. Liu, A. E. Willner, R. W. Boyd, Broadband frequency translation through time refraction in an epsilon-near-zero material. *Nature communications* **11**, 2180 (2020).

43. Z. Dong, H. Li, T. Wan, Q. Liang, Z. Yang, B. Yan, Quantum time reflection and refraction of ultracold atoms. *Nat. Photon.* **18**, 68–73 (2024).

44. A. M. Shaltout, V. M. Shalaev, M. L. Brongersma, Spatiotemporal light control with active metasurfaces. *Science* **364**, eaat3100 (2019).

45. Y. Shen, Q. Zhan, L. G. Wright, D. N. Christodoulides, F. W. Wise, A. E. Willner, K. Zou, Z. Zhao, M. A. Porras, A. Chong, others, Roadmap on spatiotemporal light fields. *Journal of Optics* **25**, 093001 (2023).

46. M. Lyubarov, Y. Lumer, A. Dikopoltsev, E. Lustig, Y. Sharabi, M. Segev, Amplified emission and lasing in photonic time crystals. *Science* **377**, 425–428 (2022).

47. T. Eichelkraut, R. Heilmann, Mobility transition from ballistic to diffusive transport in non-Hermitian lattices. *Nature communications* **4**, 2533 (2013).

48. A. Dikopoltsev, S. Weidemann, M. Kremer, A. Steinfurth, H. H. Sheinfux, A. Szameit, M. Segev, Observation of Anderson localization beyond the spectrum of the disorder. *Science Advances* **8**, eabn7769 (2022).

49. L. Xiao, X. Zhan, Z. Bian, K. Wang, X. Zhang, X. Wang, J. Li, K. Mochizuki, D. Kim, N. Kawakami, others, Observation of topological edge states in parity–time-symmetric quantum walks. *Nature Physics* **13**, 1117–1123 (2017).

50. K. Mochizuki, D. Kim, H. Obuse, Explicit definition of PT symmetry for nonunitary quantum walks with gain and loss. *Physical Review A* **93**, 062116 (2016).

51. P. Xue, X. Qiu, K. Wang, B. C. Sanders, W. Yi, Observation of dark edge states in parity-time-symmetric quantum dynamics. *National Science Review*, nwad005 (2023).

52. D. C. Brody, Biorthogonal quantum mechanics. *Journal of Physics A: Mathematical and Theoretical* **47**, 035305 (2013).

**Acknowledgments:**

**Funding:**

J.F. is supported by the Leverhulme Trust through a Study Abroad Studentship. H. P. and T. S. thank The Royal Society (grants UF160112, URF\R\221004, RGF\EA\180121 and RGF\R1\180071) (TS, HMP) and the Engineering and Physical Sciences Research Council (grant no. EP/W016141/1) A.S. acknowledges funding from the Deutsche Forschungsgemeinschaft (grants SZ 276/9-2, SZ 276/19-1, SZ 276/20-1, SZ 276/21-1, SZ 276/27-1, GRK 2676/1-2023 'Imaging of Quantum Systems', project no. 437567992, and SFB 1477 "Light-Matter Interactions at Interfaces", project no. 441234705). A.S. also acknowledges funding from the Krupp von Bohlen and Halbach Foundation as well as from the FET Open Grant EPIQUS (grant no. 899368) within the framework of the European H2020 programme for Excellent Science.

**Author contributions:**

J. F. and S. W. jointly developed the fundamental theory and performed the experiments. T. S. and H. M. P. developed the analytical theory of temporal interface dynamics and time-topological state profiles. A. S. and H. M. P. supervised the project. All authors contributed ideas, discussed the results, and co-wrote the manuscript.

**Competing interests:**

The authors declare that they have no competing interests.

**Data and materials availability:**

Data and codes used in the analysis are available on [reference to Rostock institutional repository, where these will be deposited upon publication].

**Supplementary Materials**

Materials and Methods

Supplementary Text

Figs. S1 to S4

References (47–52)

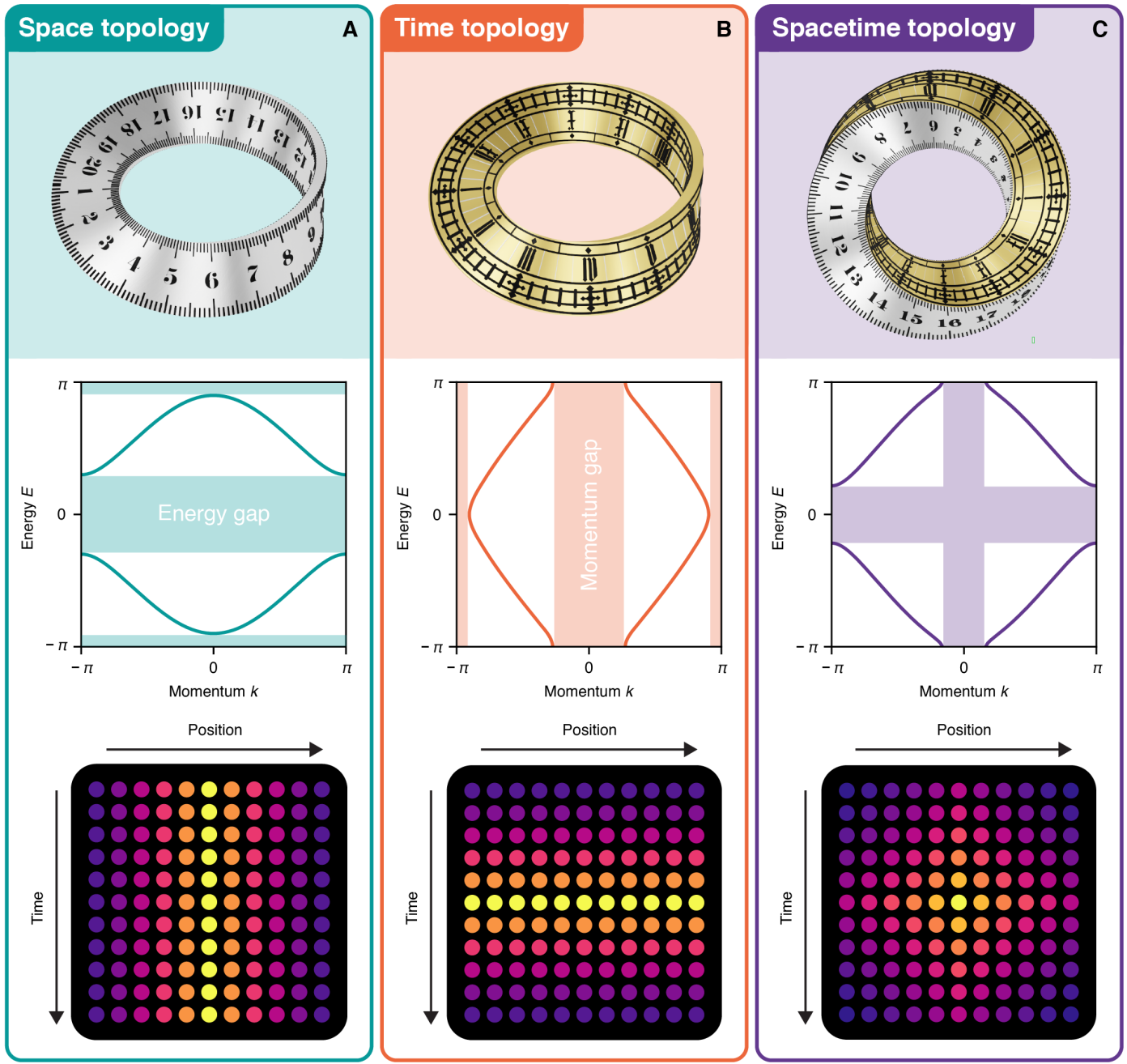

**Fig. 1. Space, time and spacetime topology. (A)** Space topology (top) relies on energy bands separated by energy gaps (center). The topology of energy gaps is associated with protected states that are localised at spatial interfaces (bottom). **(B)** Time topology (top) relies on momentum bands separated by momentum gaps (center). Time-topological states associated with the topology of momentum gaps may arise that are localised at temporal interfaces (bottom). **(C)** Spacetime topology (top) relies on bands that are gapped in energy and momentum (center). Spacetime-topological events, where topological states are localised at spacetime interfaces (bottom).

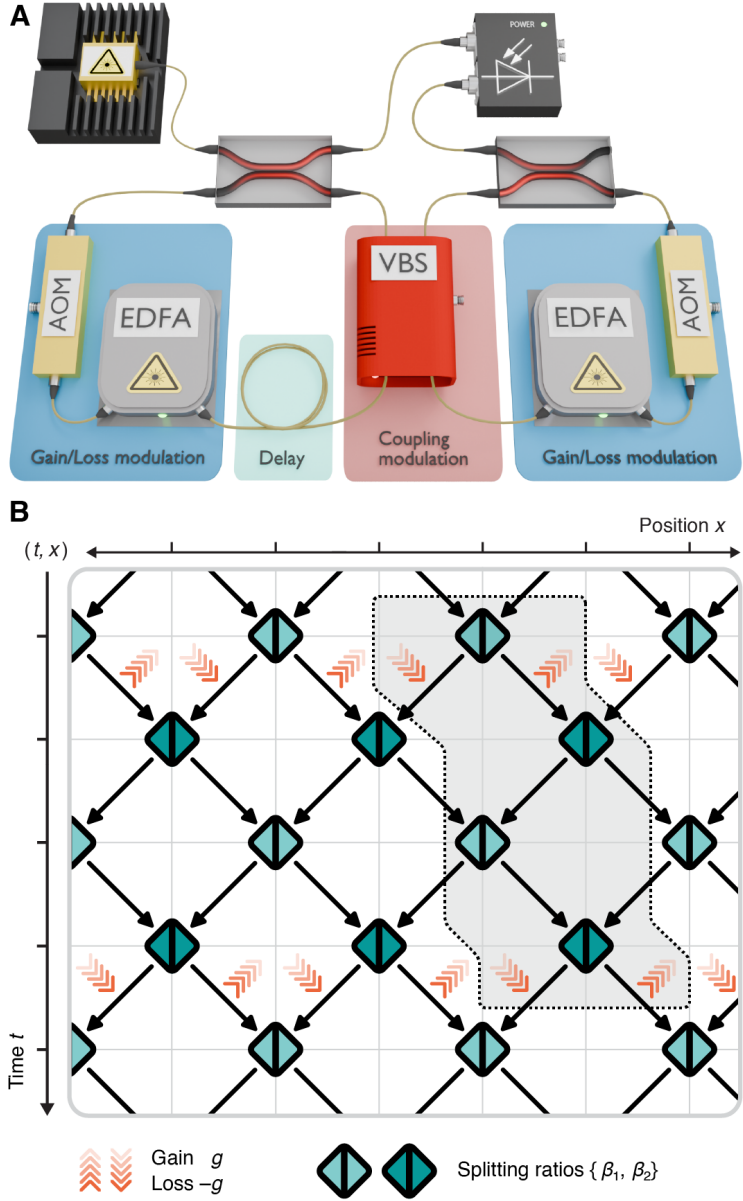

**Fig. 2. Experimental implementation of topological photonic lattices. (A)** Simplified experimental setup for the realization of photonic lattices. Two optical fiber loops are connected by a variable beamsplitter (VBS) that sets the coupling $\beta$. Acousto-optical amplitude modulators (AOMs) and erbium-doped fibre amplifiers (EDFA) enable a gain-loss modulation $\exp(\pm g)$. **(B)** The implemented photonic lattices consist of a mesh lattice of beamsplitters and gain-loss modulations with a four-time-step period, and a two-step spatial periodicity. The resulting Floquet-Bloch unit cell is shaded in grey.

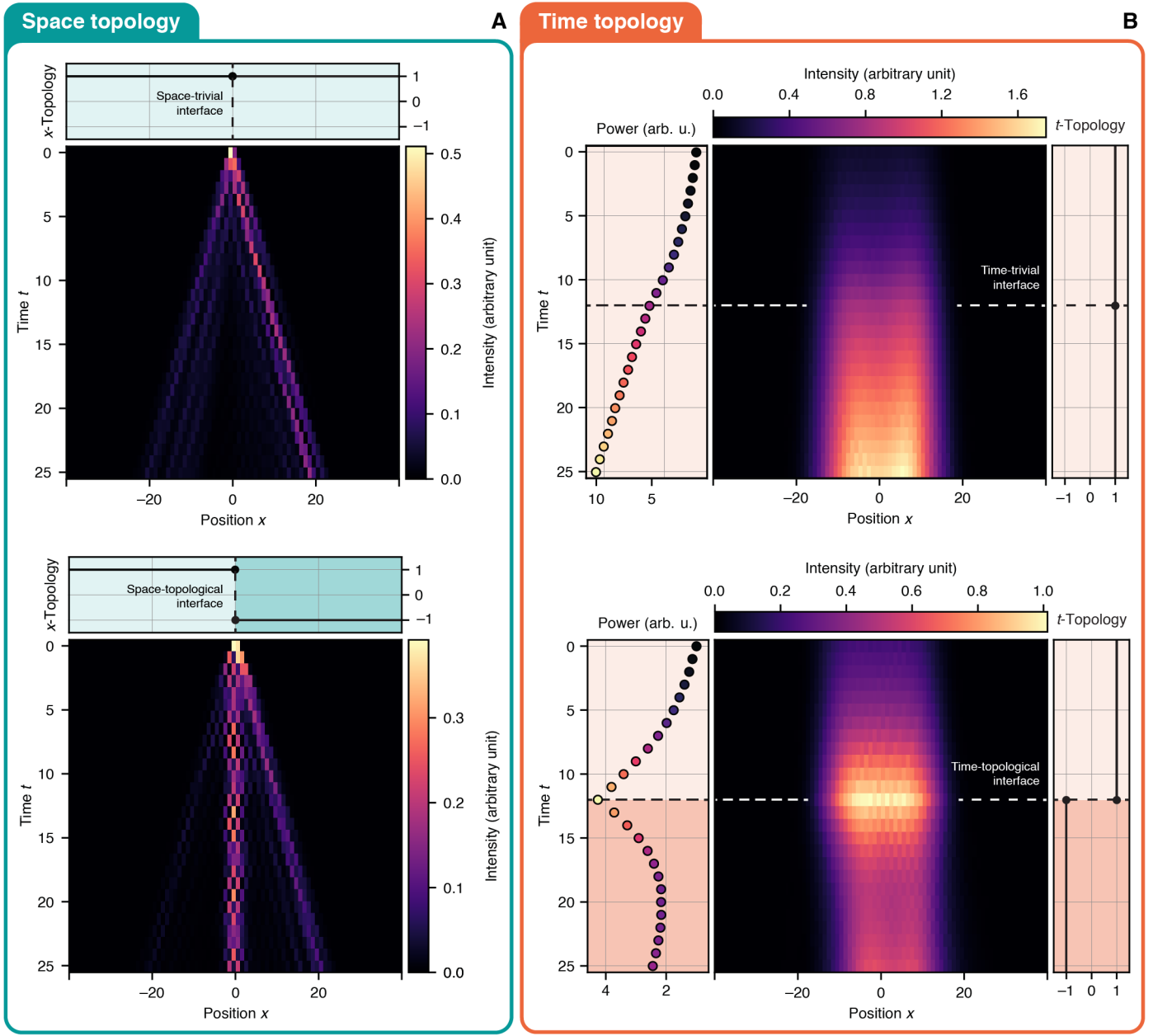

**Fig. 3. Experiments on space- and time-topological states. (A)** Light is injected at a topological (bottom) or trivial (top) spatial interface, i.e., at the junction of two lattices with different or identical space-topological invariants, respectively. The familiar spatially localised topological state emerges only at the non-trivial interface. **(B)** A spatially broad light field is injected. Here, there is no spatial interface as in **a** but instead a topological (bottom) and trivial (top) temporal interface. A time-topological state that is localised in time emerges only in the case of a non-trivial temporal interface described by time-topological invariant. The lattice parameters and calculation of the topological invariants are detailed in Section 1 and 5 of the Supplementary Text, respectively.

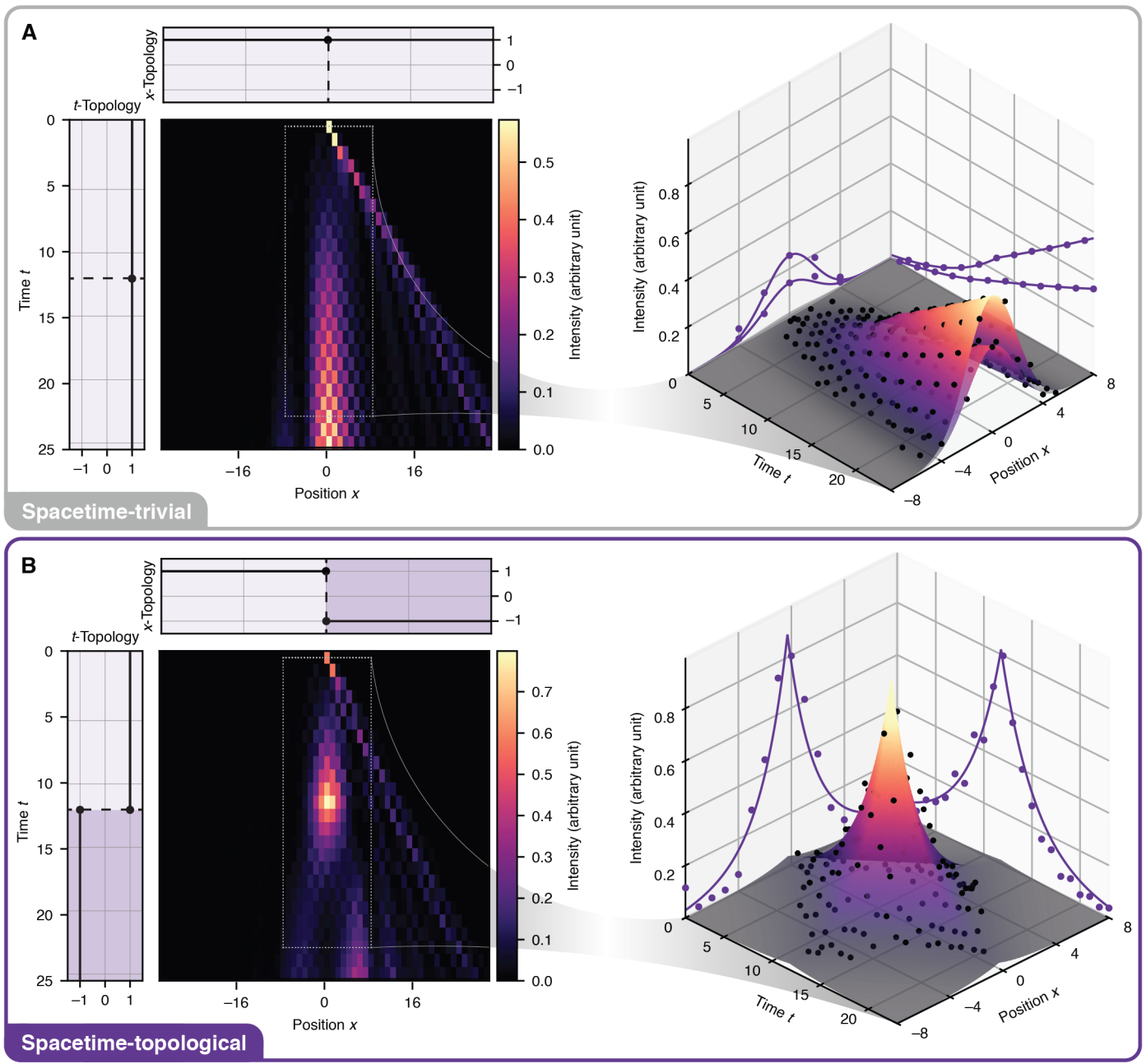

**Fig. 4. Observation of a spacetime-topological event. (A)** In the spacetime-trivial case, the light evolution shows neither exponential spatial nor any temporal localisation at the intersection of a spatial and temporal interface. This is particularly evident for the data in the proximity of the intersection, where the evolution is described well by fits of two growing modes of Gaussian shape that belong to different sublattices. **(B)** At the intersection, a spacetime-topological event, where a topological state localises at a single point in spacetime, can be seen. Light is injected at the spatial topological interface. Data in the proximity of the intersection is described well by a fit of a 2D exponential decay, which is expected for the topological event. The lattice parameters and calculation of topological invariants are detailed in Section 1 and 5 of the Supplementary Text, respectively.

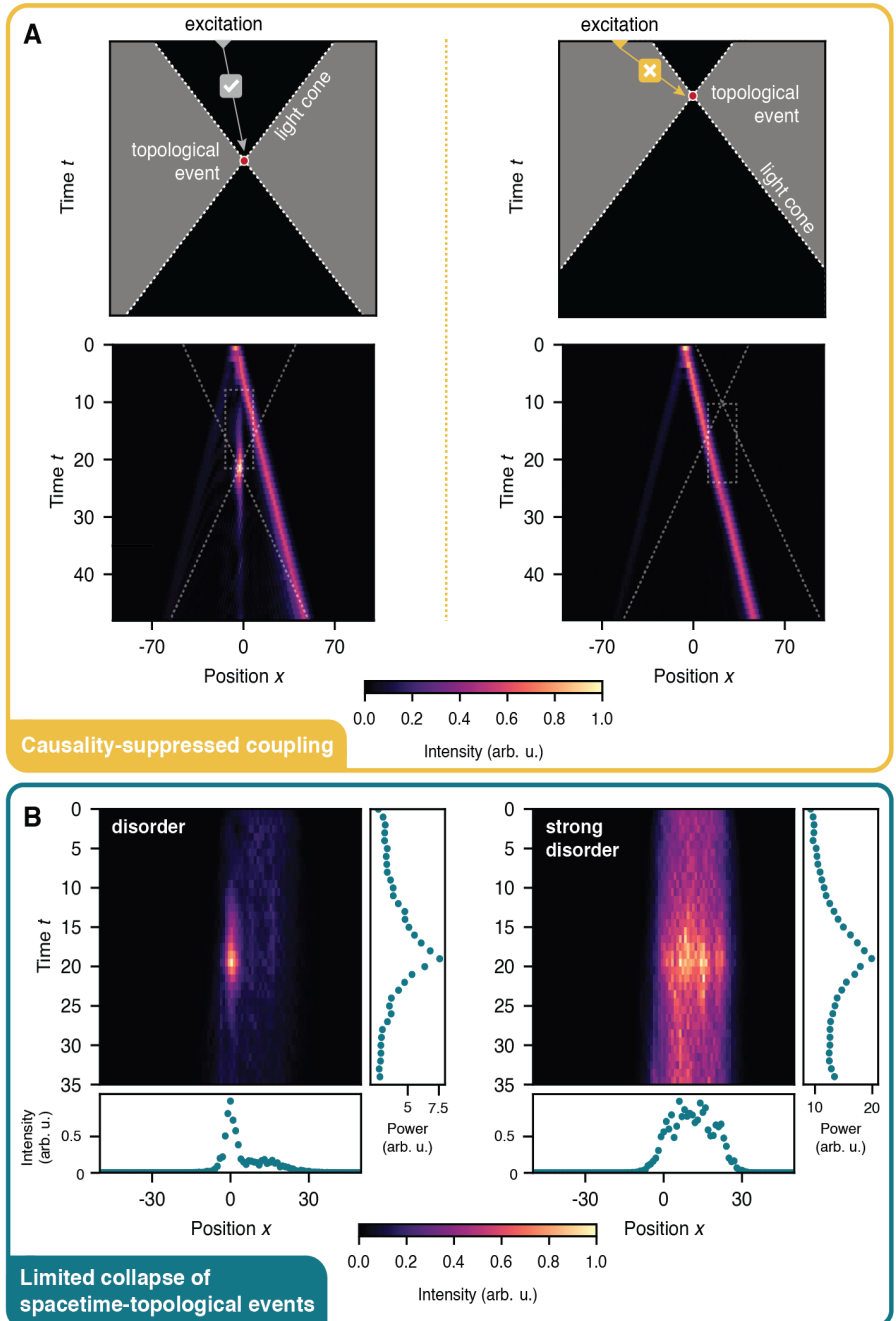

**Fig. 5. Experimental demonstration of causality-suppressed coupling and limited collapse of spacetime-topological events.** **(A)** Coupling into the spacetime-topological event is only possible from its past light cone (left), whereas coupling from its future light cone is completely suppressed due to causality (right), even for a finite overlap between excitation and the state. **(B)** Shown here is the behaviour of spacetime-topological events under phase disorder, i.e., $\varphi_u(t,x) \to \varphi_u(t,x) + \delta(x)$ with $\delta(x)$ randomly chosen from an interval $[-\Delta\varphi, \Delta\varphi]$. The results are averaged over multiple disorder realisations. For moderate disorder (left, $\Delta\varphi = 0.2\pi$), the spacetime localisation remains robust. However, for strong disorder (right, $\Delta\varphi = 0.4\pi$) the limited collapse of the localisation only in space occurs, whereas due to the unique nature of the energy-momentum gap, which may close only partially, the temporal localisation persists.